# Exponential Maps and Symmetric Transformations in Cluster-Spin System for Lattice- Ising Models

You-Gang Feng

Department of Basic Sciences, College of Science, Guizhou University, Cai-jia Guan, Guiyang, 550003 China

E-mail: ygfeng45@yahoo.com.cn

We defined exponential maps with one parameter, associated with geodesics on the parameter surface. By group theory we proposed a formula of the critical points, which is a direct sum of the Lie subalgebras at the critical temperature. We consider the self similar transformations as symmetric operations. In the opinion of symmetry we analyzed the hexagon lattice system, and got its three cluster spin states: single, double, and threefold, then its critical point is calculated. There are two cases for lattice-Ising model in thermodynamic equilibrium. In one case the periodic boundary conditions are present without the infinite self similar transformations; in another the system is in the possibility of the infinite self similar transformations without the conditions. We think the real exact critical points close to our results.

## 1. Introduction

We have set up cluster-spin Gaussian model for lattice-Ising model [1], by means of which some critical points with high accuracy can be calculated. A common feature is that in a cluster spin system the minimal fractal dimensions $D_{\min}$ require a fractional edge $n^*$, which doesn't satisfy the self similar transformations of the integer edges, so that two cases arise at the critical temperature: On the one hand the system approaches the critical point to require a fractional edge, on the other the self similar transformations allow only those integer edges. We then see that the system is forced to continuously adjust the edges in order to approach the critical point further. We also notice that the values of the fractal dimensions around $D_{\min}$ are very close to $D_{\min}$. For example, in the triangle lattice system $[1]$, by computing we get: $D_{\min} = 1.814055098$ for $n^* = 14.4955$, $D_1 = 1.814092989$ for $n_1 = 15$, $D_2 = 1.814445317$ for $n_2 = 13$, these fractals are almost equal, but the differences of the lattice numbers $P$ in a cluster are distinct: $P^* = 127.8$ for $D_{\min}$, $P_1 = 136$ for $D_1$, $P_2 = 105$ for $D_2$. With the change of edges, the adjustment of the lattice number of in each cluster is so drastic that it causes great fluctuations at the critical temperature. Thus, in the adjustment process different clusters will occur. With the same reason, we can explain fluctuations in other systems.

We are concerned about what complexity the phenomenon show? We have seen that in a reducible system a subcluster and an ordered reducible cluster have different fractal dimensions and coordination numbers. When the clusters with different sizes occur the system will not lie at the critical point, although the temperature is the same $T_c$ still, which implies that the critical temperature and the critical point are different parameters and the continuous phase transition is executed on a complicated parameter space.

## 2. Exponential maps

We have obtained a final general expression of the critical point with the minimal fractal dimension[1], by which we can conveniently calculate the critical point for lattice-Ising model. What is the meaning for the formula? We think that understanding its meaning will help us study deeply the model itself. Some parameters are often introduced in the investigating critical behavior and the parameters space is imagined as a complicated surface, on which the critical points are geometry points[2 – 7]. We know that in the renormalization group theory $K = J / k_B T$ is regarded as a parameter alone, and a parameter space related to $K$ is set up. In this paper we also consider $K$ as a parameter. For the same temperature $T_c$, different clusters have different coupling constants *J,* which means that the parameter $K$ can change at the critical temperature. $J$ in $K_c = J / k_B T_c$ is unique and corresponds to the minimum of the fractional dimension，and $J \neq 0$. The following symbols used in this paper are the same as that in [1]. For the plane square lattice, see figure 1, a reducible cluster contains two subclusters, which is of the fractal dimensions $D_{sq}$. The system consists of two subsystems: The first subsystem relates to a single state with its spin magnitude $S_{11}$ and coordination number $Z_{11}$ and fractal dimensions $D_{sq}$, which partition function is $Q_{11}$. The second subsystem to a double coupling state with its spin magnitude $S_{12}$ and coordination number $Z_{12}$ and fractal dimensions $D_{sq}^2$, which partition function is $Q_{12}$. We noticed that there must exist an exponential factor $\exp[(J_{11}/k_B T)S_{11}^2]$ in $Q_{11}$, there also should be an exponential factor $\exp[(J_{12}/k_B T)S_{12}^2]$ in $Q_{12}$. The coupling constants $J_{11}$ and $J_{12}$ belong to $S_{11}$ and $S_{12}$, respectively. Introducing two exponential maps $\gamma_1(K_1)$ and $\gamma_2(K_2)$ with the parameters, they are defined as

$$\gamma_1(K_1) = \exp(K_1 S_{11}^2), \qquad \gamma_2(K_2) = \exp(K_2 S_{12}^2) \quad , \tag{1}$$

where $K_1 = J_{11}/k_B T$, $K_2 = J_{22}/k_B T$. The definitions are meaningful for they appear in the partition functions $Q_{11}$ and $Q_{12}$. If $T \to +\infty$, then $K_1, K_2$ all tend to zero, and we define

$$\gamma_1(0) = b_1 \quad , \qquad \gamma_2(0) = b_2 \quad , \tag{2}$$

where $b_1$ and $b_2$ are two points on the parameter surface. It is clear to see that $\gamma_1(K_1)$ and $\gamma_2(K_2)$ are two curves starting from points $b_1$ and $b_2$, and their derivatives with respect to $K_1$ or $K_2$, at $K_1 = K_2 = 0$, are

$$\gamma_1'(0) = S_{11}^2 \quad , \qquad \gamma_2'(0) = S_{12}^2 \tag{3}$$

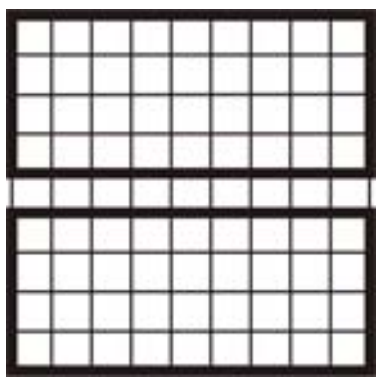

Fig.1. A reducible cluster contains two subclusters in the plane square lattice, where a subcluster is of rectangle form..

By (2) and (3), in the differential geometry sense $S_{11}^2$ and $S_{12}^2$ can be regarded as tangent vectors at points $b_1$ and $b_2$ $[8,9]$. Since $S_{11}^2$ depends on the fractal dimensions $D_{sq}$ and varies with the values of $D_{sq}$, which indicates that all values of $S_{11}^2$ together construct a vector space at the point $b_1$. Similarly, a vector space at the point $b_2$ is set up by $S_{12}^2$ with different values. The values of $K_1$ and $K_2$ increase from zero with the decreasing of temperature $T$ from infinity. Let $K_{c1} = J_{11} / k_B T_c$, which means $J_{11}$ relates to $D_{\min}$, when $K_1 = K_{c1}$, $\gamma_1(K_{c1}) = \exp(K_{c1} S_{11}^2) = C_1$, namely, the point $C_1$ be a finishing point of the curve $\gamma_1(K_1)$. The physical meaning of $K_{c1}$ is clear that it is just concerned in the critical point of the first subsystem. Suppose there were infinite curves joining the starting point $b_1$ with the moving point $C_1$ for a variety of values $S_{11}^2$. Let us now investigate the meaning of the critical point on the parameter surface. Since the length of $\gamma_1(K_1)$ is given by $[8]$

$$L_1 = \int_0^{K_{c1}} \sqrt{1 + [\gamma_1'(K_1)]^2}\, dK_1 \quad , \tag{4}$$

For the fixed intervals $[0, K_{c1}]$ and the same change of $K_1$, the integrand is a monotonic increasing function with the magnitude of $S_{11}^2$, and so the minimum of $L_1$ depends merely on $S_{11}^2$ proportional to $D_{sq,\min}$. In fact, the coupling constant $J_{11}$ in $K_{1c} = J_{11} / k_B T_c$ is certain, only related to the minimum of $S_{11}^2$, thus point $C_1$ is fixed, which uniquely to the critical point of the first subsystem. Whenever $L_1$ takes the minimum the curve $\gamma_1(K_1)$ becomes a geodesic $[8,9]$, with starting point $b_1$ and finishing point $C_1$. Meanwhile the curve $\gamma_2(K_2)$ with its arc length $L_2$ also is a geodesic from its starting point $b_2$ to the finishing point $C_2 = \gamma_2(K_{c2})$, where $K_{c2} = J_{22} / k_B T_c$, $J_{22}$ related to $D_{sq,\min}^2$, and $K_{c2}$ concerned in the critical point of the second subsystem. Since $Q_{11}$ and $Q_{12}$ exist in the system partition function [1], in a product form of $Q_{11}Q_{12}$, in which there should be a product of $\gamma_1(K_1)$ and $\gamma_2(K_2)$ as follows:

$$\gamma_2(K_2)\gamma_1(K_1) = \exp(K_2 S_{12}^2)\exp(K_1 S_{11}^2) \qquad (5)$$

In a point of view of topology, (5) is a product mapping, which means that arcs of $\gamma_1(K_1)$ and $\gamma_2(K_2)$ link up in a head-to-tail manner of addition to form a curve with the shortest lengths $L_1 + L_2$, in other words, the starting point $b_2$ of $\gamma_2(K_2)$ is just the finishing point $C_1$ of $\gamma_1(K_1)$, which shows a form of direct sum in the mathematics sense (see section 3). In the physics sense, at point $C_2$ a reducible cluster becomes simply connected and the system gets into a new hierarchy of the self similar transformations, so that the point $C_2$ should correspond to the system critical point $K_c$. The curves $\gamma_1(K_1)$ and $\gamma_2(K_2)$ with the minimal fractal dimensions $D_{sq,\min}$ can be called critical paths in the mathematics sense [9].

For the cube lattice system [1], we will get similar results. However, it should be emphasized that a double coupling state cannot present as an independent state in a reducible cluster, although its existence were possible at first sight. The reason is that: Let exponential map $\gamma_1(K_1)$ be concerned with a single state and arc length $L_1$, $\gamma_2(K_2)$ a double coupling state with arc length $L_2$, $\gamma_4(K_4)$ a fourfold coupling state with arc length $L_4$. If the double state were independent of the other states, $\gamma_2(K_2)$ would present in the partition function of the system, such that a product mapping were produced such as $\gamma_4(K_4)\gamma_2(K_2)\gamma_1(K_1)$ related to arc lengths $L_1 + L_2 + L_4$ on the parameter surface, which is longer than the arc lengths $L_1 + L_4$ to a product mapping of $\gamma_4(K_4)\gamma_1(K_1)$ without $\gamma_2(K_2)$. Obviously, the curves $L_1 + L_4$ are geodesics and $L_1 + L_2 + L_4$ are not ones, so the double state will not exist independently. With the same reason, a threefold coupling state also is impossible.

**3. Further lifting: symmetry transformations and groups**

The self similar transformations are tantamount to symmetric ones preserving the proper symmetries of the system, and those transformations to violate the original symmetries are possibly forbidden. Therefore, under such restriction a self similar transformation can simply be regarded as a result of action of a transformation group. Since there is no a symmetric operation which can change all of lattices in a cluster into a new lattice positing on the cluster center after rescaling, thus the becoming of a cluster into the new lattice is called local symmetry-breaking. Meanwhile all of new lattices still keep the original symmetry, hence a global symmetry presents still. When the system becomes ordered, which likes a single point space, the global symmetry-breaking then appears. Thus, we can describe the continuous phase transition as follows: global symmetry $\rightarrow$ local symmetry-breaking $\rightarrow$ global symmetry $\rightarrow$ local symmetry-breaking $\rightarrow \cdots \rightarrow$ global symmetry-breaking, which correspond to the process of the transformations: low hierarchy $\rightarrow$ high hierarchy $\rightarrow$ more high hierarchy $\rightarrow \cdots \rightarrow$ infinite hierarchy.

For the square lattice, see (1) and (5), the natural logarithm of $\gamma_1(K_1)$ at the critical point is $K_{c1}S_{11}^2$, and the natural logarithm of $\gamma_2(K_2)\gamma_1(K_1)$ at the critical points is $K_{c1}S_{11}^2 + K_{c2}S_{12}^2$. According to the relation of Lie groups with their algebras [9 – 11], the direct product of Lie subgroups corresponds simply to the direct sum of their subalgebras. If $\gamma_1(K_{c1}) = \exp(K_{c1}S_{11}^2)$ and $\gamma_2(K_{c2}) = \exp(K_{c2}S_{12}^2)$ are considered as two subgroups, $K_{c1}S_{11}^2$ and $K_{c2}S_{12}^2$ become their subalgebras. The rule of the self similar transformations tell us that after rescaling a single state spin $S_{11}$ becomes a lattice spin $s$ with coupling constant $j_{11}$ in the space of dimensions $D_{sq}$ related to $K_{c1}$, a double coupling state spin $S_{12}$ becomes a lattice spin $s$ with coupling constant $j_{12}$ in the space of dimensions $D_{sq}^2$ related to $K_{c2}$, hence there exist one-to-one relations of $K_{c1}S_{11}^2$ and $K_{c2}S_{12}^2$ with $(j_{11}/k_BT_c)s^2$ and $(j_{22}/k_BT_c)s^2$. Thus, we can say that $\exp[(j_{11}/k_BT_c)s^2]$ and $\exp[(j_{22}/k_BT_c)s^2]$ are subgroups corresponding to the subgroups $\exp(K_{c1}S_{11}^2)$ and $\exp(K_{c2}S_{12}^2)$. As $s^2 = 1$ and the supposition of $j_{11} = J_{11}$ and $j_{12} = J_{12}$ [1], thus $K_{c1} = j_{11}/k_BT_c$, $K_{c2} = j_{12}/k_BT_c$, and so we can say that $\exp(K_{c1}) = \exp(K_{c1}s^2)$ and $\exp(K_{c2}) = \exp(K_{c2}s^2)$ are corresponding to $\exp(K_{c1}S_{11}^2)$ and $\exp(K_{c2}S_{12}^2)$, respectively. We then obtain further lifting: for a system numbered $i$ with $k$ subsystems, in general, let there be Lie subgroups $\exp(K_{c1})$, …, $\exp(K_{ck})$ at the critical temperature, their direct product be $\exp(K_{c1})\cdots\cdots\exp(K_{ck})$ equaling a Lie group *G*. The relevant Lie algebra $K_c$ of the *G* be just the direct sum of the Lie subalgebras, which give us the formula of the critical points of the system

$$K_c = K_{c1} + \cdots + K_{ck} \quad , \tag{6}$$

where $K_{c1}$ associates with a single state spin $S_{i1}$, …, $K_{ck}$ with a $k$-fold coupling state spin $S_{ik}$ allowed for the system. The meaning of the direct sum on the parameter surface is that the curves $\gamma_1(K_1)$ and $\gamma_2(K_2)$ link up in a head-to-tail manner, their join point is $\gamma_1(K_{c1}) = \gamma_2(0)$ and the finish point is $\gamma_2(K_{c2})$, seen in section 2. Noticing the relation of the critical point with the minimal fractal dimension, we then have [1]

$$K_{c1} = 1/2D_{\min} \qquad\qquad K_{ck} = 1/2D_{\min}^k \quad , \tag{7}$$

where $D_{\min}$ is the minimal fractal dimension of a subcluster. Using (6) and (7), we can immediately get formula (43) of reference [1]. At this moment we can say that the above Lie group *G* and its subgroup $K_c$ are just the symmetry transformation groups at the critical temperature. If we know a system transformation: its clusters and the relating minimal fractal dimension, we can calculate its critical point right now. The symmetry analyses will help us find out the cluster structures.

In the following calculation of the critical point for a hexagon lattice system we try to analyze its self similar transformations from the point of view of group theory. Figure 2 illustrates a reducible cluster containing six subclusters, for simplicity, where a small triangle represents a subcluster. Figure 3 illustrates the detail structure of a sub-cluster, where a cell is a minimal hexagon containing six lattices (vertices). A small circle denotes the cell center, all circles constitute an equilateral triangle with dashed and solid lines. The total cells increase, in series of natural numbers, with increasing

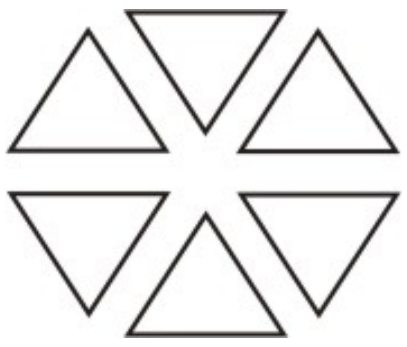

Fig.2. A reducible cluster containing six subclusters in the hexagon lattice, where a small triangle represents a subcluster in the single state.

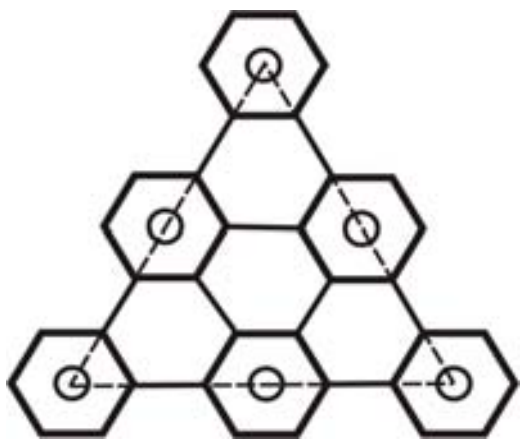

Fig. 3. The structure of a subcluster for the hexagon lattice with $n = 10$.

the triangle edge. Thus, the total lattices of the subcluster are $P = 6 \cdot (1 + 2 + \cdots)$.

Only those cells that lie on the boundary of the triangle contribute to the girth of the cluster. Let the girth of a cell be $L$, $L = 6$. A cell on a vertex of the triangle contributes five sixths of $L$ to the girth of he cluster, a cell not on the vertex one half of $L$ to the girth, a segment between two cells on the boundary one sixth of $L$. One third of the cluster girth equals the edge length of the cluster, denoted by $n$. We then have $P = 6 \cdot [1 + 2 + \cdots + (n+2)/4]$, further, by the formula (1) of the reference [1], the fractal dimension of the subcluster is defined as

$$D_{he} = \frac{Ln\{6 \cdot [1 + 2 + \cdots + (n+2)/4]\}}{Ln(n)} \quad , \tag{8}$$

where $n = 6,10,14,18,...$. Calculating (8) yields

$$D_{he,\min} = 1.54189 \quad , \qquad n^* = 14.3 \tag{9}$$

Figures 2 tell us that the subclusters keep completely the original symmetries of the system so that the system will execute only the self similar transformations of the subclusters without having the reducible cluster be ordered. For the infinite hierarchies, however, the reducible cluster will be ordered, otherwise the system has no phase transition. Now we are facing a question: how much multi-fold coupling states are there in a reducible cluster for the infinite transformations? Before answer this question, let us investigate the relation between a multi-fold coupling state and the symmetry of the system.

Symbols of symmetric operations used in the follows come from [12]. See figure 1, exchanging two subcluster spin states in a double coupling state of the square lattice will not alter the coupling state, which shows some symmetries of the square. For example, the symmetric operations $S_2$ and $C_{2v}$. Similarly, see figure 4, the exchanging two single states in a fourfold coupling state of a cube lattice leaves the fourfold coupling state unchanged and the operation agrees with some symmetries of the cube. For example, $C_4$, $C_{2v}$ and $S_2$. For a fourfold coupling state in the reducible cluster of the cube lattice, analyzing dimensionality we find: $D_{cu}^4 = D_{cu} \cdot D_{cu} \cdot D_{cu} \cdot D_{cu} = D_{cu}^2 \cdot D_{cu}^2 = D_{cu} \cdot D_{cu}^3$, where $D_{cu}$ corresponds to a single state, $D_{cu}^2$ to a double coupling state, $D_{cu}^3$ to a threefold coupling state. A single state is essential, without it there is no subcluster. The threefold coupling state doesn’t satisfy the symmetric properties of the cube, so it cannot exist. The above equality means that a fourfold coupling state might equivalently be regarded as a coupling state of two double states. Whenever a double coupling state appears a fourfold coupling state presents simultaneously, because of the nearest neighbor interactions and the symmetry of the cube. Thus, we say that the double coupling states are only involved in a fourfold coupling state, they are not independent of the fourfold coupling state, which coincides with the conclusion given by section 2. Now let us consider a hexagon lattice. Figure 5, 6, and 7 demonstrate three double coupling states of dumbbell forms, two threefold coupling states of triangle forms, and one six-fold coupling state of a hexagon, respectively. In the figures, a segment between two subclusters represents both of interaction and correlation length, and a small circle represents both of a single state and a subcluster, respectively. Occurrences of threefold coupling states and double coupling states only depend on the next-nearest neighbor interactions. In the symmetry sense, these states satisfy some symmetries of the hexagon. For example, there is a symmetric operation $C_{3v}$ in a threefold coupling state (figure 6). The symmetric operation $S_2$ exists in a double coupling state (figure 5). In the figure 7, however, a six-fold coupling state is forbidden actually. The reason is that if the state arises, the reducible cluster turns a new lattice, see figure 2, all of which then form a triangle lattice system instead of a hexagon lattice one to violate the original symmetries. How can a reducible cluster become ordered without the six-fold coupling state? Thinking of a case in which there is a “final” reducible cluster on a hierarchy, and the next-nearest neighbor and the nearest neighbor change into indistinguishable, such that three double coupling states and two threefold coupling states can exist. From figures 5 and 6, we see that two threefold coupling states are not connected to each other, three double coupling states play bridge roles in joining one threefold coupling state to another, and so the whole reducible cluster becomes a simply connected domain. Furthermore, the centers of these subclusters with double coupling state or threefold coupling state are the same, which means the interactions of them are the nearest. Thus, six single states, three double coupling states, and two threefold coupling states independently lie on the same hierarchy. If we relate the “final” reducible cluster to the correlation length, the case will be achievable provided the correlation length reaches infinity. Making use of (6), (7) and (9) we then get the critical point of the hexagon lattice system

$$K_c = \frac{1}{2D_{he,\min}} + \frac{1}{2D_{he,\min}^2} + \frac{1}{2D_{he,\min}^3} = 0.6703 \qquad (10)$$

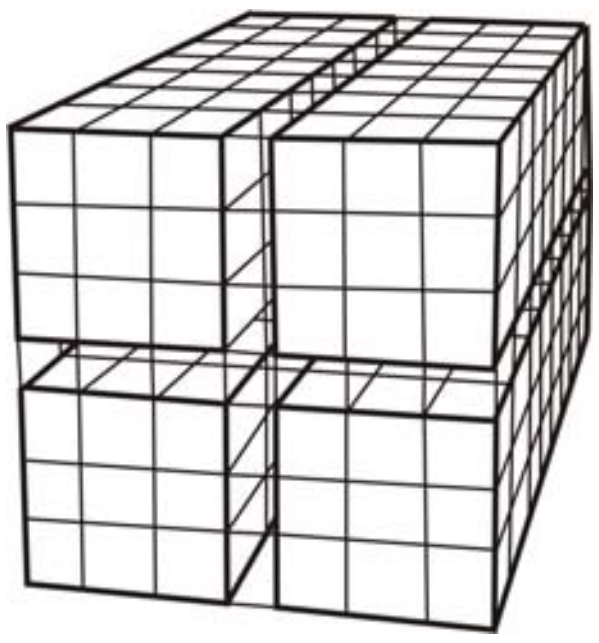

Fig. 4. A reducible cluster with four subclusters for the cube lattice, where a subcluster is of a cuboid form.

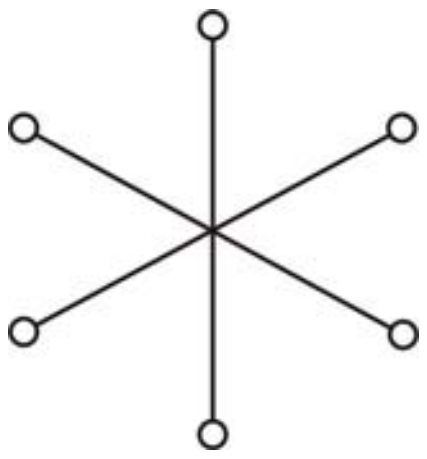

Fig. 5. Three double coupling states of dumbbell forms.

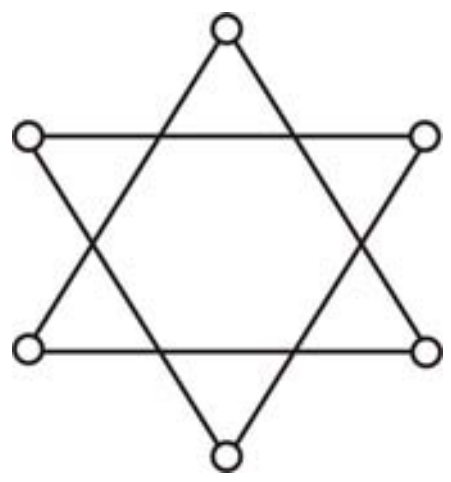

Fig. 6. Two threefold coupling states of triangle forms.

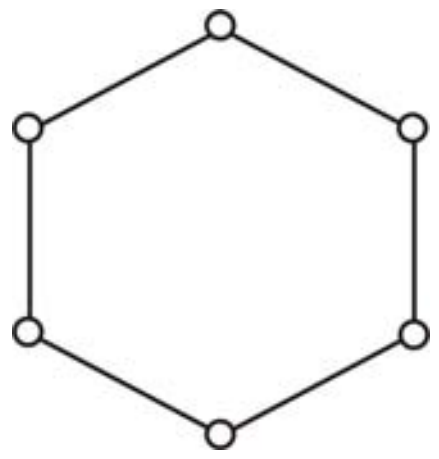

Fig. 7. One six-fold coupling state of a hexagon form.

The critical point given by duality transformations with the help of periodic boundary conditions is 0.6585 [6]. The duality transformations require integer edges. If $n = 14$, then $D_{he} = 1.55144$, and inserting the $D_{he}$ to (10) instead of $D_{he,\min}$ related to $n^* = 14.3$, we get $K_c = 0.6639$, which is very close to 0.6585.

**4. Conclusion**

We defined exponential maps with one parameter, which are associated with geodesics on the parameter surface. From the point of view of group theory we proposed a formula of the critical points, which is a direct sum of the Lie subalgebras at the critical temperature. We consider the self similar transformations as symmetric operations. In the opinion of symmetry, we analyzed the hexagon lattice system and obtained its three cluster spin states: single state, double coupling state and threefold coupling state, then its critical point is obtained. Up to now, we have seen that there are two cases for lattice-Ising model in thermodynamic equilibrium. In one case the periodic boundary conditions are present without the infinite self similar transformations; in another the system is under the necessity of the self similar transformations of infinite hierarchies without the periodic boundary conditions. Two cases correspond to two different sets of the critical points, which is reasonable? According to the Ergodic hypothesis both cases can exist, but the probability of existence of the case without the periodic boundary conditions is far larger than that with the conditions. Thus, we think that the reasonable magnitudes of the critical points should be close to our results.